\documentclass{PoS}

\title{VHE $\gamma$-ray/X-ray correlation studies in Mrk 421 down to the quiescent state} 

\ShortTitle{VHE $\gamma$-ray/X-ray correlation in Mrk 421}

\author{\speaker{Barbara Patricelli}\\
        Universit\`a di Pisa, INFN - Sezione di Pisa, Instituto de Astronom\'ia - UNAM\\
        E-mail: \email{barbara.patricelli@pi.infn.it}}

\author{Magdalena Gonz\'alez\\
        Instituto de Astronom\'ia - UNAM\\
        E-mail: \email{magda@astro.unam.mx}}
        
\author{Nissim Fraija\\
        Instituto de Astronom\'ia -UNAM\\
        E-mail: \email{nifraija@astro.unam.mx}}        

\abstract{The blazar Mrk 421 is one of the closest, brightest and fastest varying source in the extragalactic X-ray/TeV sky. In the last years, many multi-wavelength campaigns have been carried out to study the correlation between the very high energy (VHE) $\gamma$-ray and X-ray fluxes of this source and, although the activity in these two energy ranges seems to be correlated in many observations, no conclusive results have been achieved yet. In this work we present a robust study of the VHE $\gamma$-ray/X-ray correlation of Mrk 421 with data taken with different VHE experiments on different time scales and for different levels of activity of the source, with special focus on the low activity states. In particular, we discuss the robustness of the correlation at the lowest fluxes corresponding to the quiescent state of Mrk 421.}

\FullConference{The 34th International Cosmic Ray Conference,\\
		30 July- 6 August, 2015\\
		The Hague, The Netherlands}

\begin{document}

\section{Introduction}

Blazars are active galactic nuclei (AGN) with relativistic jets closely aligned to our line of sight. They show intense flux variability on different time scales and at all wavelengths. Their broadband spectral energy distribution (SED) shows a double-peaked structure, with a ``hump'' at low energies, peaking in radio through X-rays and a second ``hump'' at higher energies, peaking in $\gamma$-rays. While there is ample evidence that the low energy peak is due to synchotrotron radiation from the relativistic electrons within the jet, there are two competing scenarios to explain the high energy peak: the so-called leptonic models (e.g. synchrotron self-Compton, SSC or external Compton, EC) and the hadronic models.  To put some constraints on the proposed theoretical models, simultaneous multiwavelenght observations of these sources over long periods of times are needed. For instance, a useful tool to discriminate between the models is the study of the correlation between VHE $\gamma$-rays and X-rays, since such a correlation is expected in the SSC scenario, while it is difficult to be explained within hadronic models (see, however, \cite{2003APh....18..593M,2013MNRAS.434.2684M}). 

In the last years, several studies have been performed to investigate the VHE $\gamma$-ray/X-ray correlation in different blazars. Positive heterogeneous correlations between the VHE $\gamma$-ray and the X-ray emission have been reported during flaring activity of several blazars (see, e.g., \cite{2008ApJ...677..906F,2006ApJ...646...61G}), although there are also cases in which the correlation is fairly loose (see, e.g. \cite{2006ApJ...641..740R}), or even  ``orphan TeV flares'', i.e. VHE $\gamma$-ray flares without X-ray activity have been reported (see \cite{2005ApJ...630..130B}). 
Positive correlations have been found also when considering long term observation periods covering different levels of activity of the source, from the low state to the highest states. For instance, Acciari et al. 2014 \cite{2014APh....54....1A} reported on a 14-year monitoring of the blazar Mrk 421 and they showed that a correlation between the X-ray and TeV energy bands is present, although the found correlation presents a considerable scatter.                                                                                   

The low state has been poorly studied at VHE, mainly due to the limited sensitivity of the current VHE $\gamma$-ray observatories and because the VHE $\gamma$-ray observations are usually only triggered when the source is in a high state of activity at other wavelengths. There are, however, a few studies pointing out the presence of a correlation also for the quiescent state of blazars. For instance, Aleksi\'c et al. 2015 \cite{2015A&A...576A.126A} recently reported on a positive correlation between the VHE $\gamma$-ray and X-ray emission of the blazar Mrk 421 during a low state of activity, and suggested that the physical processes dominating the emission during non-flaring states have similarities with those occurring during flaring activity for this source.


A direct comparison of all these results would be useful to better investigate the possible differences among flares and between the flaring and the quiescent state of blazars, as well as to study if the correlation depends on the time scale; however, this is a difficult task because data from different experiments are reported in an heterogeneous way (i.e., for VHE $\gamma$-ray data different energy thresholds and units are used). 

Here we combine together several sets of simultaneous VHE $\gamma$-ray and X-ray observations taken by different experiments, on different time scales and for different monitoring periods for the blazar Mrk 421. We convert all the data to the same energy threshold/range and units, to make them directly comparable and present a study of the correlation between the X-ray and the VHE $\gamma$-ray emission. In particular, our analysis is focused on  the lower fluxes characterizing the quiescent state of Mrk 421.

\section{The data sets}\label{sec:data}

The most unbiased and comprehensive data set of simultaneous VHE $\gamma$-ray/X-ray observations of Mrk 421 is the one presented by Acciari et al. 2014  \cite{2014APh....54....1A}, that reported on a long term (14-year) monitoring of Mrk 421 with the Whipple 10 m telescope. They combined the Whipple data with simultaneous RXTE/ASM data and found that the VHE $\gamma$-ray fluxes are linearly correlated with the X-ray fluxes on monthly time-scale, although the present large scatter was not quantified. To investigate the possible dependence of the correlation from the time scale and from the experiment by which data have been taken, we combine the Whipple-RXTE/ASM data with other published data sets that allow a good flux sampling over a large range of flux values (from the quiescent state flux up to the most intense flare fluxes) and over different time scales. We then apply a statistical approach that allow us to estimate if all the data are correlated, the robustness of the correlation and how significant are the deviations from it. In the following we briefly summarize the characteristics of the additional data sets we used.

\emph{The HEGRA CT1-RXTE/ASM data. -} Aharonian et al. 2003 \cite{2003A&A...410..813A} reported on the monitoring of Mrk 421 with HEGRA CT1 during a period of intense activity of the source in 2001. They combined the HEGRA CT1 data with simultaneous RXTE/ASM data and found that the VHE $\gamma$-ray fluxes are linearly correlated with the X-ray fluxes on hourly time scale.

\emph{The MAGIC/Whipple/VERITAS-XMM Newton data. -} Acciari et al. 2009 \cite{2009ApJ...703..169A} reported on the monitoring of Mrk 421 during two flaring states: one on April 2006, with VHE $\gamma$-ray data from MAGIC and Whipple and, the other on May 2008 with data from VERITAS. They combined the VHE $\gamma$-ray data with simultaneous XMM-Newton data, in the energy range of 0.5-10 keV (similar to RXTE/ASM data) and pointed out the lack of correlation on a time scale of $\sim$ 20-30 minutes during the first flare.

\emph{The MAGIC-RXTE/ASM data. -} Albert et al. 2007 \cite{2007ApJ...663..125A} reported on the monitoring of Mrk 421 from November 2004 to April 2005 with MAGIC. They combined the MAGIC data with simultaneous RXTE/ASM data and found that the VHE $\gamma$-ray fluxes are correlated with the X-ray fluxes on a nightly time scale. 

To compare all these data sets, the measured VHE $\gamma$-ray and X-ray fluxes must be converted to a common energy threshold (or energy range) and be expressed in the same units. For the VHE $\gamma$-ray data we choose an energy threshold of 400 GeV and the Crab units, while for the X-ray data an energy range (2-10 keV) and the units of RXTE/ASM counts/s, the same reported in \cite{2014APh....54....1A}. 

To convert the VHE $\gamma$-ray fluxes from an energy threshold $E_0$ to the new energy threshold $E_1$ (400 GeV) in Crab units we use the following relation:
\begin{eqnarray}
F(E > E_1)=F(E > E_0)\times \frac{\int_{E_1}^{\infty} \phi(E) dE} {\int_{E_0}^{\infty}\phi(E) dE \times \int_{E_1}^{\infty}\phi_{\rm Crab}(E) dE},
\end{eqnarray}
where $\phi(E)$ and $\phi_{\rm Crab}(E)$ are the source and the Crab spectrum respectively. To do this conversion, we take the spectra observed by the corresponding instrument and when possible for the same time period of the corresponding observation to minimize discrepancies between different data sets due to systematic effects of the experiment \cite{2003A&A...410..813A,2009ApJ...703..169A,2007ApJ...663..125A,2004ApJ...614..897A,2006APh....25..391H,2008ICRC....2..691G,2008ApJ...674.1037A}.

To convert the X-ray fluxes in RXTE/ASM rates we use the online WebPIMMS\footnote{http://heasarc.gsfc.nasa.gov/cgi-bin/Tools/w3pimms/w3pimms.pl} tool.

\section{The VHE $\gamma$-ray/X-ray correlation}

We study the VHE $\gamma$-ray/X-ray correlation for Mrk 421 using all the data sets described in the previous section, with all the fluxes converted to the chosen units and energy threshold/range. 

The VHE $\gamma$-ray and X-ray fluxes are linearly correlated, with a Pearson correlation coefficient $R=0.81$. To estimate how robust is the correlation we use the maximum likelihood approach described in \cite{2005physics..11182D}. This method assumes that, if the correlated data (in our case the X-ray and VHE $\gamma$-ray fluxes $F_{x,i}$ and $F_{\gamma,i}$) can be described by a linear function $F_{\gamma}=a F_{x}+b$ with an intrinsic scatter $\sigma$, the optimal values of the parameters ($a$, $b$ and $\sigma$) can be determined by minimizing the minus-log-likelihood function:
\begin{eqnarray}
L(a,b,\sigma) =  \frac{1}{2} \sum_i \log(\sigma^2+\sigma_{F_{\gamma,i}}^2+a^2 \sigma_{F_{x,i}}^2)+ \frac{1}{2} \sum_i \frac{(F_{\gamma,i}-a F_{x,i}-b)^2}{\sigma^2+\sigma_{F_{\gamma,i}}^2+a^2 \sigma_{F_{x,i}}^2},
\end{eqnarray}
where $\sigma_{F_{x,i}}$ and $\sigma_{F_{\gamma,i}}$ are the uncertainties on $F_{x,i}$ and $F_{\gamma,i}$ respectively. 

\begin{figure}[h!]
\begin{center}
\includegraphics[scale=0.8]{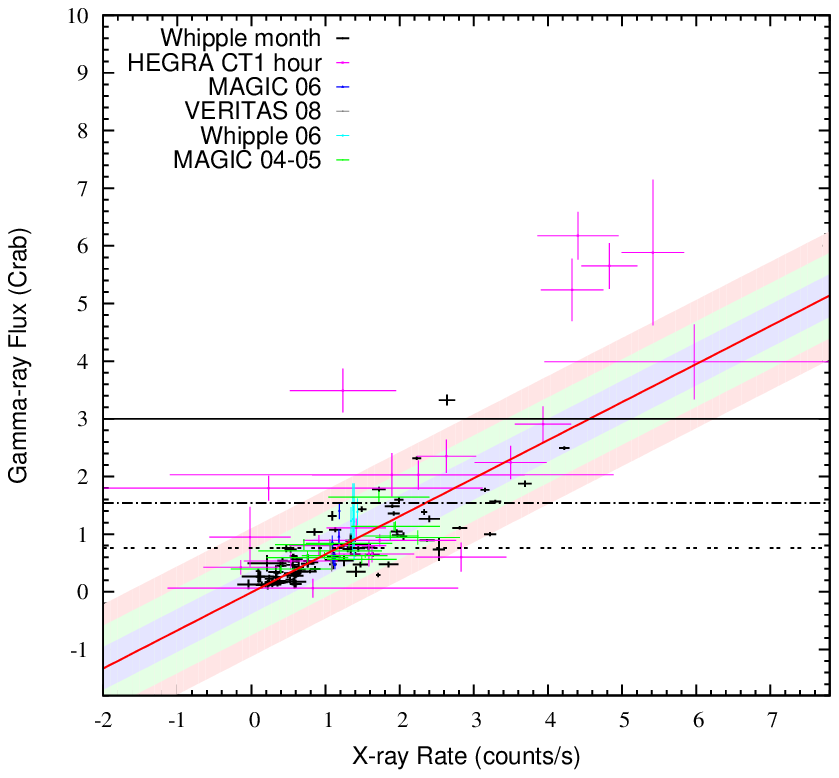}\includegraphics[scale=0.8]{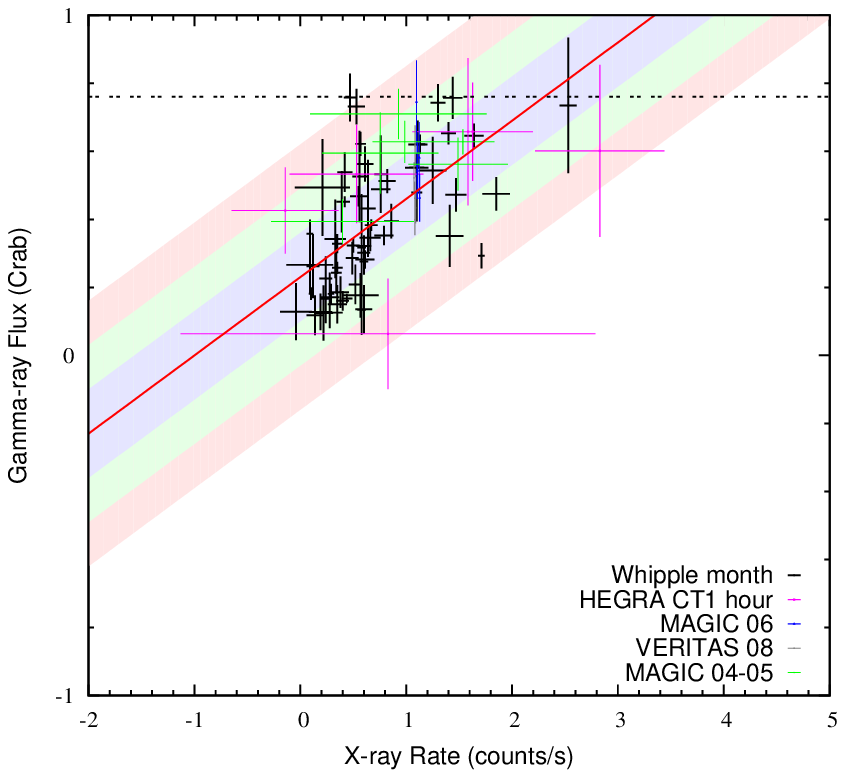}
\caption{Left panel: correlation between the X-ray (2-10 keV) and the VHE $\gamma$-ray (E $>$ 400 GeV, Crab) fluxes of Mrk 421 with data from different X-ray satellites and VHE $\gamma$-ray experiments (see text). The red solid line is the best fitting straight line. The shadowed regions represent the 1, 2 and 3 $\sigma$ scatter around the best fit of this correlation; for comparison, the ``very low'', ``low'' and ``high'' state flux levels of Mrk 421, as measured by VERITAS \cite{2011ApJ...738...25A} are shown as black dotted, dot-dashed and solid lines respectively. Right panel: same as left panel, but the fit has been done considering only data with VHE $\gamma$-ray fluxes up to 0.76 Crab, corresponding to the VERITAS ``very low'' state flux of Mrk 421 \cite{2011ApJ...738...25A}.}\label{fig:allcorr}
\end{center}
\end{figure}

With this method we obtain $a= 0.66 \pm 0.04$, $b=-0.01 \pm 0.05$ and $\sigma=0.37 \pm 0.03$. The results are shown in the left panel of Fig. \ref{fig:allcorr}. It can be seen that almost all the data are positively correlated within 3 $\sigma$, therefore the correlation is robust and consistent between instruments. Such a correlation is naturally expected within the SSC scenario. However, there are intense VHE $\gamma$-ray fluxes that seem not to be accompanied by intense X-ray fluxes, lying at more than 3 $\sigma$ above the best linear fit. It is interesting to note that these fluxes are all higher than the ``high state'' flux\footnote{We refer as the VERITAS ``high'' state flux the value corresponding to the ``high state A'' reported in Acciari et al. 2011 \cite{2011ApJ...738...25A}.} of Mrk 421 (corresponding to 3 Crab) as measured by VERITAS \cite{2011ApJ...738...25A}. Another interesting point is that the dispersion around the best fit straight line is smaller in the region of low VHE $\gamma$-ray fluxes, where the majority of the points lie within 1-2 $\sigma$. This is an unexpected result: in fact, due to the limited sensitivity of the detectors, lower fluxes are expected to have larger uncertainties and larger dispersion. 

To better study this last aspect, we repeat the whole procedure only for the quiescent state of Mrk 421. 
From all the data sets we select the points with a VHE $\gamma$-ray flux up to the VERITAS  ``very low'' state flux of Mrk 421 \cite{2011ApJ...738...25A}, corresponding to 0.76 Crab\footnote{To convert the VERITAS fluxes in Crab units, we assume the VERITAS spectrum of the Crab as reported by Holder et al. 2006 \cite{2006APh....25..391H}.} for an energy threshold of 400 GeV. 
This value is comparable with the upper limit on the quiescent flux of Mrk 421 reported by Tluczykont et al. 2010 \cite{2010A&A...524A..48T} (0.33 Crab for an energy threshold of 1 TeV, corresponding to $\approx$ 0.73 Crab above 400 GeV\footnote{For the conversion to the energy threshold of 400 GeV we assume the VERITAS spectra for the Crab and Mrk 421 \cite{2006APh....25..391H,2011ApJ...738...25A}.}): this means that the observed fluxes below the VERITAS ``very low'' state are dominated by the emission of the source in the quiescent state. 

These data are linearly correlated, with $R=0.61$. By fitting them with the maximum likelihood approach we obtain $a= 0.23 \pm 0.03$, $b=0.23 \pm 0.03$ and $\sigma=0.13 \pm 0.01$. The results are shown in the right panel of Fig. \ref{fig:allcorr}. It can be seen that all the data lie within 3 $\sigma$ from the best fit straight line. Therefore, also when considering the quiescent state independently, the correlation is robust, consistently with the results found by Aleksi\'c et al. 2015 \cite{2015A&A...576A.126A}: this suggests that SSC dominates the emission of Mrk 421 not only during flares, but also during the quiescent state. It is interesting to note that the intrinsic scatter of the correlation for the quiescent state is about a factor of 3 smaller that the one obtained for the overall correlation. 

The correlation for the quiescent state and the one obtained fitting together all the data are consistent, within 3 $\sigma$, up to X-ray fluxes of $\sim$ 4 counts/s.

\section{Conclusions}
We presented a study of the VHE $\gamma$-ray/X-ray correlation for the blazar Mrk 421. We  used several data collected by different VHE $\gamma$-ray experiments and X-ray satellites on different time scales and for different monitoring periods of the source. We converted all the fluxes in the same units and with the same energy threshold/range, in order to make them directly comparable and we performed a statistical study of the overall correlation, as well as of the correlation for the quiescent state of the source. We found that in both cases data are linearly correlated and that the correlations are robust and consistent within instruments,  suggesting SSC as the dominating emission process. We also found that the two correlations are consistent between themselves within 3 $\sigma$, and that the dispersion of the data around the best fit straight line is smaller for the quiescent state. 
A more detailed analysis of the results here presented and of their theoretical interpretation will be presented elsewhere. 

\acknowledgments
This work was supported by Luc Binette scholarship and DGAPA-UNAM under PAPIIT project IG100414-3.

\end{document}